# Leveraging symmetry for an accurate spin-orbit torques characterization in ferrimagnetic insulators

Martín Testa-Anta[1,*], Charles-Henri Lambert[2], Can Onur Avci[1,*]

[1]*Institut de Ciència de Materials de Barcelona (ICMAB-CSIC), Campus de la UAB, 08193 Bellaterra, Spain*

[2]*Department of Materials, ETH Zürich, Hönggerbergring 64, CH-8093 Zürich, Switzerland*

## Abstract

Spin-orbit torques (SOTs) have emerged as an efficient means to electrically control the magnetization in ferromagnetic heterostructures. Lately, an increasing attention has been devoted to SOTs in heavy metal (HM)/magnetic insulator (MI) bilayers owing to their tunable magnetic properties and insulating nature. Quantitative characterization of SOTs in HM/MI heterostructures are, thus, vital for fundamental understanding of charge-spin interrelations and designing novel devices. However, the accurate determination of SOTs in MIs have been limited so far due to small electrical signal outputs and dominant spurious thermoelectric effects caused by Joule heating. Here, we report a simple methodology based on harmonic Hall voltage detection and macrospin simulations to accurately quantify the damping-like and field-like SOTs, and thermoelectric contributions separately in MI-based systems. Experiments on the archetypical Bi-doped YIG/Pt heterostructure using the developed method yield precise values for the field-like and damping-like SOTs, reaching -0.14 and -0.15 mT per $1.7\times10^{11}$ A/m$^2$, respectively. We further reveal that current-induced Joule heating changes the spin transparency at the interface, reducing the spin Hall magnetoresistance and damping-like SOT, simultaneously. These results and the devised method can be beneficial for fundamental understanding of SOTs in MI-based heterostructures and designing new devices where accurate knowledge of SOTs is necessary.

## I. INTRODUCTION

Spin-orbit torques (SOTs) are the generic name given to the current-induced torques in ferro-/ferrimagnetic heterostructures with large spin-orbit coupling and broken inversion symmetry mainly driven by, but not limited to, bulk spin Hall (SHE) and interfacial Rashba-Edelstein effects.[1] During the past decade, SOTs have become the state-of-the-art magnetic manipulation method by electrical currents and enabled magnetization switching,[2,3,4] domain wall and skyrmion motion,[5,6,7] steady-state magnetic oscillations,[8,9] and magnon generation/suppression[10,11] in convenient device geometries. These experimental achievements have given rise to a multitude of spintronics device concepts with memory, logic, signal transmission and computing functionalities suitable for the complementary metal-oxide-semiconductor (CMOS) industry and the post-CMOS computing era.[12,13,14,15] SOTs have been originally discovered and extensively studied in all-conducting heavy metal (HM)/ferromagnetic (FM) bilayers such as Pt/Co, Ta/CoFeB and W/CoFeB.[16,17,18,19,20] However, the research has rapidly expanded into other materials such as antiferromagnets,[21,22,23] ferrimagnets,[24,25,26] topological insulators[27,28,29] and magnetic insulators (MIs)[4,30,31], among others.

Recently, the interest in MIs in the SOTs context has rapidly grown.[5,32,33,34] MIs offer many advantages over their conducting counterparts thanks to their tunable magnetization and anisotropy, low Gilbert damping[35], and long spin diffusion lengths[36]. These attributes, together with their insulating nature leading to reduced Ohmic losses when integrated into microelectronic circuits, make MIs a highly attractive material platform for low power magnonic and spintronic applications.[37,38,39] Within the family of MIs, yttrium iron garnet ($Y_3Fe_5O1_2$, YIG) is particularly appealing due to its record low Gilbert damping (~$10^{-4}$ - $10^{-5}$), making it ideal for nano-oscillators,[40,41] and magnonic devices,[42,43,44,45] with SOT-enabled operation. Despite the vital importance of SOTs in the integration of MIs in potential spintronic concepts, their accurate and straightforward characterization by simple electrical methods is intrinsically difficult. Harmonic Hall voltage (HHV)[4,16,17,46] and spin-torque ferromagnetic resonance[47] measurements are the two established methods in this context but rely on the Hall resistance and magnetoresistance output signals, respectively, which are orders of magnitude smaller in HM/MI systems with respect to all-metallic systems. The few existing studies have considered only the damping-like (DL)-SOT characterization,[4,46,48,49,50] but the separate quantification of the field-like (FL)-SOT, equally important for the SOT-driven magnetization dynamics,[51,52,53,54,55,56] and a proper account of current-induced Joule heating in these measurements and interfacial spin transport properties remained elusive thus far.

In this article, we show a simple method for an accurate SOTs quantification in MIs relying on the combination of HHV measurements and macrospin simulations. We develop, and test with simulations, a HHV measurement scheme using a non-standard geometry, which allows us to exploit simple symmetry arguments to disentangle the effective fields due to the damping-like ($B_{DL}$) and field-like ($B_{FL}$) SOTs, and thermoelectric contributions precisely. The quantification of both SOTs components and thermoelectric effects yield an intrinsic theoretical error as low as ~0.2% on the simulated data. As a proof-of-concept, we apply the proposed approach on a Bi-doped YIG/Pt bilayer with in-plane (IP) magnetic anisotropy obtaining $B_{DL}$ = -0.15 mT and $B_{FL}$ = -0.14 mT and a thermoelectric field of $E$ = 0.27 V/m per $j$ = 1.7×10$^{11}$ A/m$^2$ injected current, all values in the range expected of such systems. Finally, current-dependent measurements show that Joule heating reduces the spin mixing conductance ($G^{\uparrow\downarrow}$) at the HM/MI and results in a systematic reduction of the spin Hall magnetoresistance (SMR) and $B_{DL}$ up to ~15%, simultaneously.

## II. EXPERIMENTAL DETAILS

**Sample preparation**. An ~18-nm-thick Bi-doped YIG (Bi:YIG from hereon) layer was rf-sputtered from a stoichiometric target onto a single-crystal Sc-substituted gadolinium gallium garnet substrate ($Gd_3Sc_2Ga_3O_{12}$, GSGG) with a base pressure <5×10$^{-8}$ Torr. The growth temperature and the Ar partial pressure during the Bi:YIG deposition were 800 ℃ and 1.5 mTorr, respectively. After the deposition, the samples were annealed for 30 min at the deposition temperature and subsequently cooled down to room temperature in vacuum. In order to have a clean interface, a 4 nm-thick Pt was dc-sputtered in-situ at room temperature without breaking the vacuum. The Ar pressure during the Pt deposition was 3 mTorr. After the Pt deposition, the continuous Bi:YIG/Pt layers were patterned into Hall bar structures by means of standard photolithography followed by top-down ion milling. The lateral dimensions of the Hall bars were 10 μm (current line width) × 30 μm (distance between two Hall crosses).

**Hall effect measurements**. HHV measurements were performed by applying an AC voltage along the current line modulated at $\omega/2\pi$ = 1092 Hz and simultaneously measuring the first and second harmonic Hall voltage responses with a lock-in amplifier. The current amplitude through the device under test was determined by connecting a 10 Ω resistor in series and reading the voltage drop across with a digital multimeter before or during the measurements. The acquired HHVs were converted into Hall resistances using the relation $R_\omega^H = V_\omega^H/I$ for comparability with the literature. For the field scans, an out-of-plane (OOP) or in-plane (IP) DC magnetic field was swept in the $\pm$600 mT or $\pm$200 mT range, respectively, while keeping the azimuthal angle ($\varphi$) fixed. For the polar ($\theta$) angle scans, the sample was rotated in the presence of a constant DC field in the 0º-360º range using a motorized rotation stage at an angular speed of 3 degrees/second. All measurements were performed at room temperature and all the data presented were averaged over five scans to improve the signal-to-noise ratio, unless specified otherwise.

## III. RESULTS AND DISCUSSION

### A. Structural and magnetic characterization

A structural characterization of the as-deposited Bi:YIG/Pt heterostructure was carried out by means of high-resolution X-ray diffraction (XRD). Figure 1a shows a symmetric $\theta$-$2\theta$ scan around the (444) diffraction peak of the GSGG substrate, occurring at 50.44º. The data do not show a clear emergence of the Bragg reflection for the Bi:YIG phase, but only a small shoulder at the right side of the substrate peak. Such observation may be attributed to a reduced crystallinity of the magnetic film or, more likely in our case, to the overlap between the substrate and Bi:YIG (444) reflections. Indeed, in the absence of interfacial strain, Bi-substitution leads to an increase in the cubic lattice parameter, resulting in a lower $2\theta$ position compared to the bulk YIG (expected at 51.09º)[57]. Additionally, the fact that the Bi:YIG peak is not shifted to lower $2\theta$ values with respect to the substrate indicates that: *i)* its OOP lattice parameter is smaller than that of GSGG, and *ii)* the tetragonal distortion induced by the lattice mismatch with the substrate is not significant.[58] Owing to the negative (111) magnetostriction constant for this particular iron garnet,[59] this absence of moderate interfacial strain will favor the occurrence of IP magnetic anisotropy. X-ray reflectivity (XRR) analysis was performed in order to check the thickness of Bi:YIG confirming the anticipated value of 18 nm. The sample topography was addressed through atomic force microscopy (AFM) before and after the Pt deposition. Figure 1b is a representative AFM image acquired from the Bi:YIG/Pt sample. The sample surface is rather flat with an RMS roughness of 0.66$\pm$0.13 nm, similar to the roughness before Pt deposition (<1 nm, not shown). This number was estimated by averaging measurements of five different regions on the sample, which highlights the overall good quality of the films.

The IP hysteresis loop recorded at 300 K using the superconducting quantum interference device (SQUID) magnetometry is displayed in Figure 1c. The large remanence accompanied by a negligible coercivity (<1 mT) are characteristic of IP magnetized garnets due to their small magnetocrystalline anisotropy. Preferential IP magnetization of Bi:YIG was also corroborated by magneto-optical Kerr effect (MOKE) measurements as shown in Figure 1d. Data recorded in polar (signal proportional to the OOP component of the magnetization) and longitudinal (signal proportional to the IP component of the magnetization) MOKE configurations with corresponding field sweeps clearly show that the IP and OOP are the easy and hard axes, respectively. Additional longitudinal MOKE measurements at different azimuthal angles showed a negligible difference (data not shown), hence, the sample is hereafter assumed to exhibit easy-plane anisotropy.

Notably, the experimental value of the saturation magnetization ($M_S$~85 emu/cm$^3$) lies well below that of the bulk YIG (~140 emu/cm$^3$ at 300 K).[60,61] Such reduction can be explained considering the $Bi^{3+} \leftrightarrow Y^{3+}$ substitution at the dodecahedral sites, as this will expand the crystal lattice and weaken the superexchange interactions between the $Fe^{3+}$ cations located at the tetrahedral and octahedral sites, primarily responsible for the ferrimagnetic behavior in iron garnets.[62] Another possibility relates to an oxygen off-stoichiometry, occurring due to the annealing at very low $O_2$ pressures. Under these conditions, the formation of oxygen vacancies has been reported to proceed via $Fe^{3+}$ to $Fe^{2+}$ reduction, giving rise to a decrease in magnetization.[63] Nevertheless, our results are consistent with literature values for Bi:YIG thin films grown on GGG or GSGG substrates,[62,64] and exploring the underlying cause of the reduced magnetization is beyond the scope of the present study.

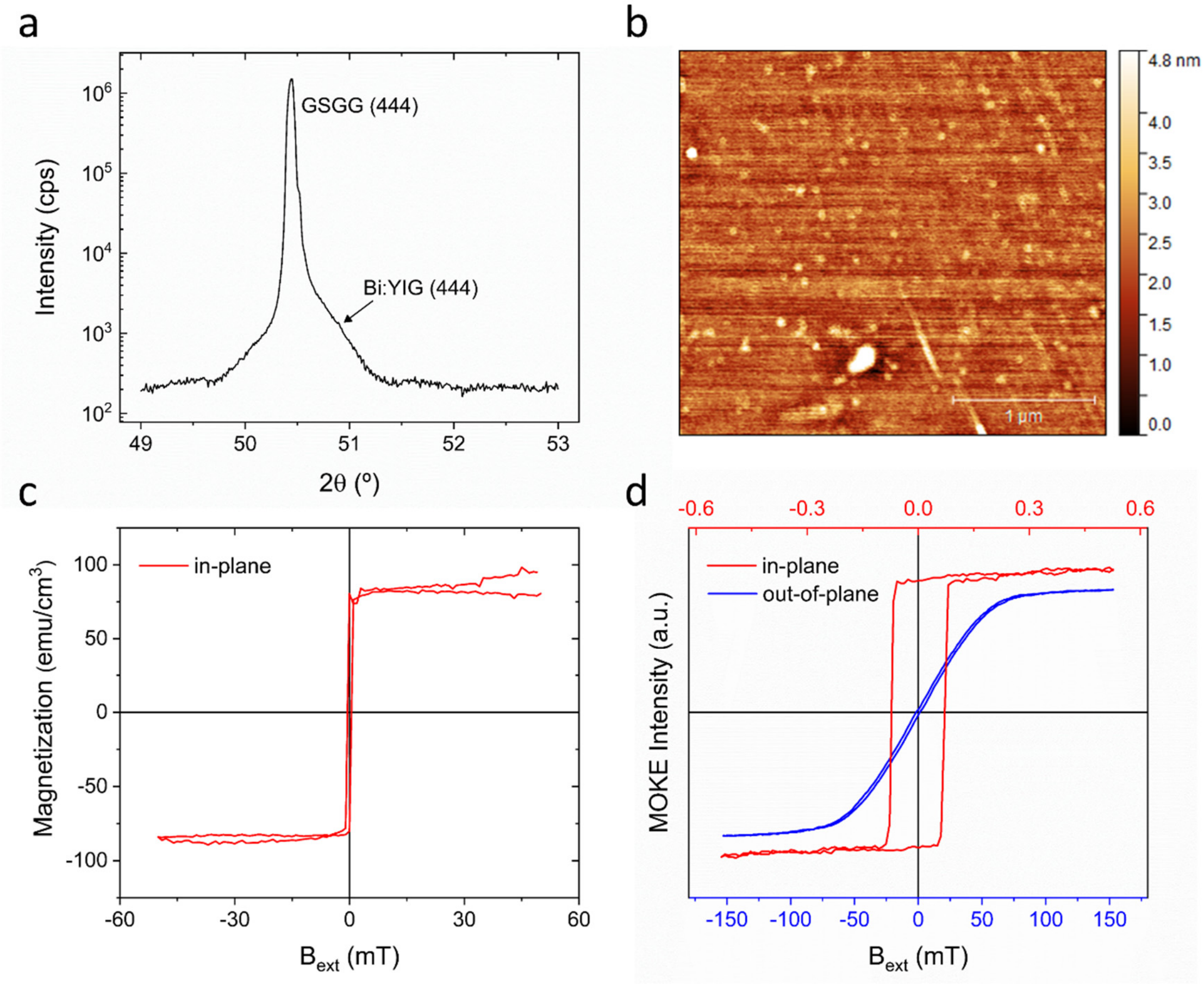

**Figure 1.** Structural and magnetic characterization of the Bi:YIG/Pt sample: (a) $\theta$-$2\theta$ scan of the Bi:YIG (18 nm)/Pt (4 nm) bilayer deposited onto a (111)-oriented GSGG substrate; (b) Representative AFM image of the previous film, revealing a rms roughness ($S_q$) value of 0.66$\pm$0.13 nm (averaged over different regions in the sample); (c) In-plane magnetization as a function of an external magnetic field (after subtraction of the linear paramagnetic background), obtained by SQUID magnetometry; (d) MOKE response as a function of an in-plane (red, top axis) or out-of-plane (blue, bottom axis) magnetic field in their respective measurement geometries (i.e., longitudinal and polar), which confirms the in-plane anisotropy of the prepared sample.

## B. First harmonic Hall characterization

The magnetotransport properties of Bi:YIG/Pt were inferred through SMR measurements and reported in Figure 2. In Figure 2a, a current injection through the Pt layer along the x-axis produces a pure spin current due to the SHE and other possible charge-spin conversion mechanisms with polarization along the y-axis and flowing along the z-axis reaching Bi:YIG. The absorption and reflection of this spin current at the Bi:YIG/Pt interface modulates the longitudinal and transverse resistance of Pt depending on the relative orientation between the magnetization and spin accumulation, which lies at the origin of SMR in HM/MI bilayers.[65] The SMR manifests itself on the Hall voltage (which we mainly use in this study) with signals proportional to both IP ($\propto m_x m_y$) and OOP ($\propto m_z$) components of the magnetization.[65,66] Therefore, these two SMR contributions exhibit the very same symmetry as the planar and anomalous Hall resistances in conducting magnetic materials, respectively, but with typical values several orders of magnitude lower.

According to the angle definitions given in Figure 2a, the Hall resistance at the first harmonic of the current frequency, $R_\omega^H$, can be expressed as follows:

$$R_\omega^H = R_{SMR}\sin^2\theta_M \sin(2\varphi_M) + R_{SMR-AHE}\cos\theta_M + R_{OHE}B_{ext}\cos\theta_B \quad (1)$$

where $R_{SMR}$, $R_{SMR,AHE}$ and $R_{OHE}$ represent the resistance modulation due to SMR, SMR-induced anomalous Hall effect (SMR-AHE) and the ordinary Hall effect (OHE), respectively. $B_{ext}$ and $\theta_B$ are the magnitude and polar angle of the external magnetic field whereas $\theta_M$ and $\varphi_M$ are the polar and azimuthal angles of the magnetization vector. Note that the OHE is the property of the Pt itself and has, in principle, no relationship with the magnetic layer underneath.

Figure 2b shows the first harmonic Hall resistance in a swept OOP field. Besides the linear OHE contribution, the signal difference at positive and negative field at high amplitudes reflect the magnetization vector switching between the up and down states. By symmetry operations with respect to $B_z$ = 0, we find that the magnetization saturates along the z-axis at around $B_z$ = 45 mT, which we denote as $B_{sat}$. Since the Bi:YIG layer possesses in-plane anisotropy, this value is equivalent to the sum of the demagnetizing field and the effective perpendicular anisotropy field possibly due to the reminiscent OOP anisotropy driven by a small lattice mismatch undetected in the XRD measurements. From the experimental data we find $R_{SMR-AHE}$ = -1.36 mΩ for an rms current density of $1.7\times10^{11}$ A/m$^2$, which yields $\rho_{SMR-AHE}$ = $-5.4\times10^{-12}$ Ω·m. A symmetric component is likewise visible in the OOP field-scan, responsible for the inverse U-shape signal and spikes observed at low fields. This additional contribution is ascribed to the reorientation of the magnetization vector within the film plane for $B_z$<$B_{sat}$ and possibly domain formation at very small fields leading to extra contributions in $R_\omega^H$ due to the non-zero $m_x m_y$ components.

IP field sweep measurements at different azimuthal angles ($\varphi_B$) are depicted in Figure 2c. The modulation of $R_\omega^H$ due to SMR shows a maximum (minimum) at $\varphi_B$ = 45º (135º), consistent with the positive $R_{SMR}$ coefficient in YIG/Pt.[4] The $\varphi_B$-dependence of the data is displayed in Figure 2d. The fitting following eq. 1 reveals a $R_{SMR}$ value of 11.23 mΩ, corresponding to $\rho_{SMR}$ = 4.5×10$^{-11}$ Ω·m. Note that we assume $\varphi_B \approx \varphi_M$ due to easy-plane anisotropy, which is further corroborated by the similar saturation fields observed for all IP field-scans. Our results show therefore that $R_{SMR}$ is much larger than $R_{SMR-AHE}$ (a factor 8.25), as customary in MI/Pt bilayers.

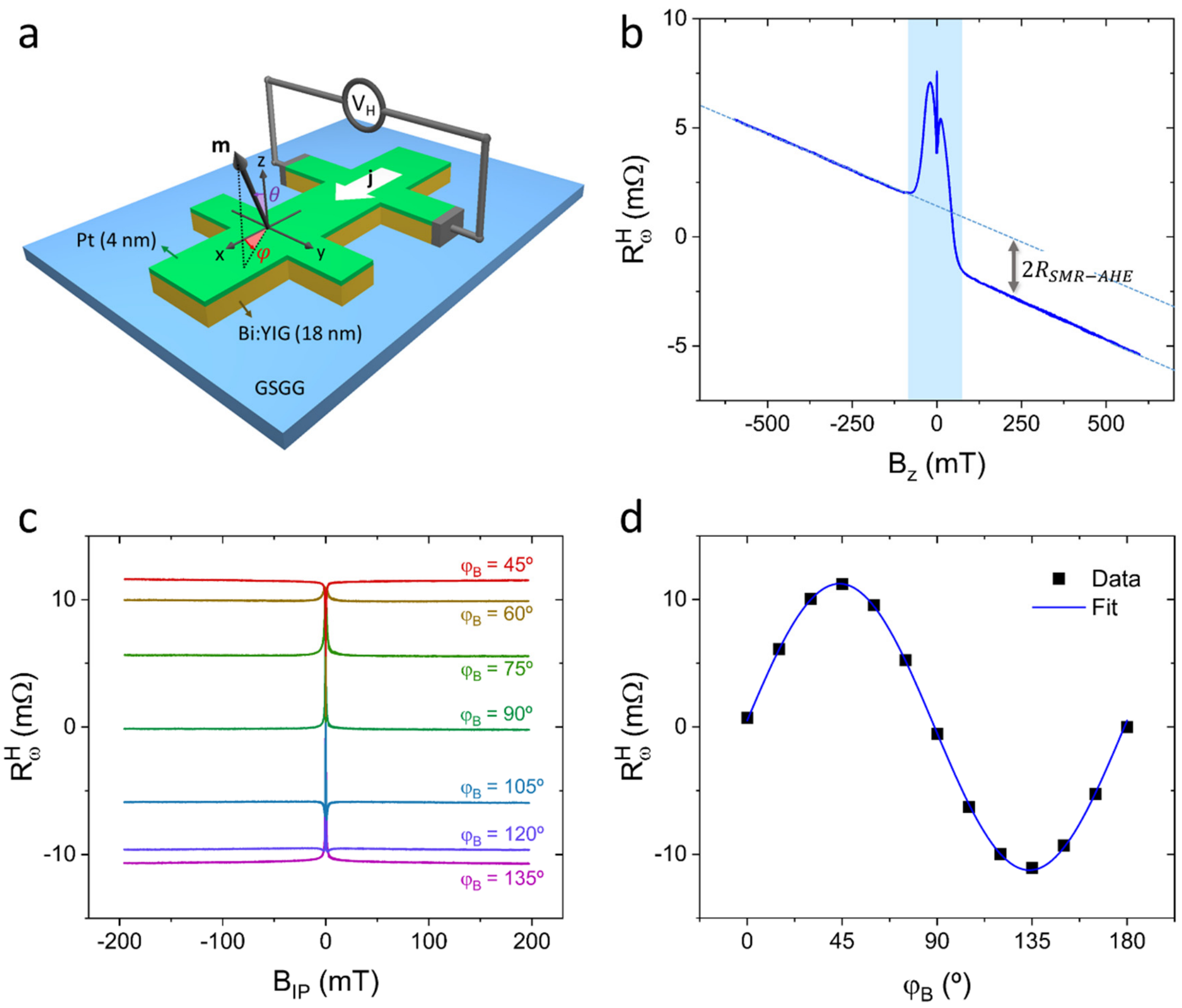

**Figure 2.** Harmonic Hall characterization of the Bi:YIG/Pt sample: (a) Schematic representation of the device geometry and corresponding coordinate system; Transverse Hall resistance measured upon sweeping an external (b) OOP or (c) IP (at different $\varphi_B$ angles) magnetic field; (d) $\varphi_B$-dependence of the transverse resistance measurements summarized in **c**, fitted to a sin2$\varphi$ function in accordance with eq. 1. All measurements were conducted at a current density (rms) of $j$ = 1.7×10$^{11}$ A/m$^2$. Note that a constant offset has been subtracted from the raw data shown in **b** and **c**.

## C. Description of the SOTs and macrospin model

Macrospin simulations are a useful tool not only to verify that the experimental data can be reproduced with existing theoretical models, but also to ascertain the symmetry of the different contributions to the current-induced fields. It is established that SOTs consist of two distinct components acting on the magnetization such as a field-like torque with symmetry $\mathbf{T_{FL}} \propto \mathbf{m} \times \mathbf{y}$, and a damping-like torque defined as $\mathbf{T_{DL}} \propto \mathbf{m} \times (\mathbf{y} \times \mathbf{m})$. Here, $\mathbf{m}$ stands for the unit magnetization vector and $\mathbf{y}$ the in-plane axis transverse to the current flow (see Figure 2a). The effect of the $\mathbf{T_{FL}}$ is therefore equivalent to an in-plane field $\mathbf{B_{FL}}$ acting along $\mathbf{y}$, whereas that of $\mathbf{T_{DL}}$ is an effective field $\mathbf{B_{DL}}$ perpendicular to the magnetization, with rotational symmetry within the $zx$-plane. Upon applying an AC current, these current-induced SOTs will induce periodic oscillations to the magnetization about its equilibrium position at the same AC frequency, dictated by the balance between the demagnetizing ($\mathbf{B_{dem}}$), anisotropy ($\mathbf{B_{ani}}$) and external ($\mathbf{B_{ext}}$) fields. In a macrospin approximation, we can then write the total torque ($\mathbf{T_{tot}}$) as:

$$\mathbf{T_{tot}} = \mathbf{T_{dem}} + \mathbf{T_{ani}} + \mathbf{T_{ext}} + \mathbf{T_{FL}} + \mathbf{T_{DL}}$$
$$= M_s(\mathbf{m} \times \mathbf{B_{dem}} + \mathbf{m} \times \mathbf{B_{ani}} + \mathbf{m} \times \mathbf{B_{ext}} - B_{FL}\mathbf{m} \times \mathbf{y} - B_{DL}\mathbf{m} \times \mathbf{y} \times \mathbf{m}) \quad (2)$$

In the present work, we simulated the equivalent first and second harmonic signals using the Scilab open source software for numerical computations.[67] More specifically, we performed a polar angle scan of the external field ($\theta_B$) in the 0º-360º range beyond the OOP saturation field of the magnetization, applied at a fixed azimuthal angle $\varphi_B$. The equilibrium configuration ($\theta_M$, $\varphi_M$) is calculated for each set of ($\theta_B$, $\varphi_B$, $\mathbf{B_{ext}}$) by minimizing eq. 2. Subsequently, the Hall resistance is computed according to the following relations:

$$R_{I+}^{H} = R_{SMR}\sin^2\theta_M\sin2\varphi_M + R_{SMR-AHE}\cos\theta_M + r_{SSE}\sin\theta_M\cos\varphi_M \quad (3)$$

$$R_{I-}^{H} = -R_{SMR}\sin^2\theta_M\sin2\varphi_M - R_{SMR-AHE}\cos\theta_M + r_{SSE}\sin\theta_M\cos\varphi_M \quad (4)$$

where $R_{I+}^{H}$ and $R_{I-}^{H}$ constitute the Hall resistances corresponding to a positive and negative DC current, respectively. Here we note that reversing the current direction changes the sign of the Hall effect coefficients but not the last term related to the current-induced thermoelectric contribution, which remains constant irrespective of the current direction. The unavoidable occurrence of Joule heating primarily creates a vertical temperature gradient, owing to the large thermal conductivity differences between the substrate and the air surrounding the device (see Figure 3a, right panel). This will create spin Seebeck effect (SSE) in the MI[68] and subsequently an inverse SHE voltage in Pt, whose contribution to the Hall resistance is quantified through the parameter $r_{SSE}$ and added to eq. 3 and 4. We then find the equivalent first and second harmonic resistances by applying the following operations to the above computed signals:

$$R_{\omega}^{H} = \frac{1}{2}(R_{I+}^{H} - R_{I-}^{H}) \quad (5)$$

$$R_{2\omega}^{H} = \frac{1}{2}(R_{I+}^{H} + R_{I-}^{H}) \quad (6)$$

## D. Simulations of the second harmonic response

Based on the model detailed above, we first discuss the harmonic signals occurring during $\theta$-scans at $\varphi_B$ = 90º, which is the standard geometry for quantifying SOTs in MI-based heterostructures.[4,17] The simulations were carried out using the experimental values of $B_{sat}$, $R_{SMR}$ and $R_{SMR-AHE}$ (previously indicated in Section III-B), and, for symmetry comparison, the individual DL-SOT, FL-SOT and SSE contributions are separately depicted in Figures 3b-d (black traces). In this geometry, the SSE contribution vanishes since it is proportional to $\cos\varphi_M$, making thus $\varphi_B$ = 90º an ideal geometry for the SOTs quantification. In this configuration $B_{FL}$ ($B_{DL}$) drives the OOP (IP) magnetization oscillations, such that they scale with $R_{SMR-AHE}$ ($R_{SMR}$), respectively (see Figure 3a). This justifies the much smaller contribution of $B_{FL}$ to the second harmonic response, which is about one order of magnitude lower than that of $B_{DL}$ even though the same field amplitudes are inputted. In terms of symmetry, they both display a similar $\theta_B$-dependence as explained below.

In the limit of small oscillations of the magnetization and assuming a linear relationship between the field and the current, the second harmonic resistance (for an angle-scan) can be expressed as follows:[17]

$$R_{2\omega}^{H} = [R_{SMR-AHE} - 2R_{SMR}\cos\theta_M\sin(2\varphi_M)]\frac{d\cos\theta_M}{d\theta_B}\frac{B_I^{\theta}}{\cos(\theta_B-\theta_M)B_{eff}} + R_{SMR}\sin^2\theta_M\frac{d\sin(2\varphi_M)}{d\varphi_B}\frac{B_I^{\varphi}}{\sin\theta_B\cos(\varphi_B-\varphi_M)B_{ext}} + r_{SSE}\sin\theta_M\cos\varphi_M \quad (7)$$

where $B_I^{\theta}$ ($B_I^{\varphi}$) represents the component of the current-induced field ($B_I$) that induces a change in $\theta_M$ ($\varphi_M$). The first term in eq. 7 accounts for the OOP oscillations and will be counteracted by the demagnetizing field, hence the effective field is defined as $B_{eff} = B_{ext} + B_{dem}$. Note that, due to the negligible anisotropy field of the Bi:YIG sample, $B_{dem}$ is herein approximated as $\approx B_{sat}$.

During the simulation of the $\theta$-scans, we set $B_{ext} = 120$ mT, which is beyond the OOP saturation field of 45 mT. Thus, the magnetization can be assumed to be saturated along $B_{ext}$ throughout the entire angle-scan ($\theta_M \approx \theta_B$). Recalling as well that the sample displays easy-plane anisotropy ($\varphi_M \approx \varphi_B$), in the vicinity of $\varphi_B$ = 90º eq. 7 reads:

$$R_{2\omega}^{H} = R_{SMR-AHE}\frac{d\cos\theta_M}{d\theta_M}\frac{B_I^{\theta}}{B_{ext}+B_{dem}} + R_{SMR}\sin\theta_M\frac{d\sin(2\varphi_M)}{d\varphi_M}\frac{B_I^{\varphi}}{B_{ext}} \quad (8)$$

Since $\mathbf{B_{FL}} = B_{FL}\mathbf{y}$ and $\mathbf{B_{DL}} = B_{DL}\mathbf{y}\times\mathbf{m}$, in the framework of a spherical coordinate system (defined by unit vectors $\mathbf{e_r}$, $\mathbf{e_\theta}$ and $\mathbf{e_\varphi}$) $B_I^{\theta}$ and $B_I^{\varphi}$ are given by:

$$B_I^{\theta} = \cos\theta_M\sin\varphi_M B_{FL} + \cos\varphi_M B_{DL} \quad (9)$$

$$B_I^{\varphi} = \cos\varphi_M B_{FL} - \cos\theta_M\sin\varphi_M B_{DL} \quad (10)$$

Again, for an azimuthal angle $\varphi_B$ = 90º one obtains $B_I^{\theta}$ = $\cos\theta_M B_{FL}$ and $B_I^{\varphi}$ = $-\cos\theta_M B_{DL}$. Taking these relations into account and computing the derivative terms in eq. 8, it follows that:

$$R_{2\omega}^{H} = -\frac{1}{2} R_{SMR-AHE} \sin(2\theta_M) \frac{B_{FL+Oe}}{B_{ext} + B_{dem}} + R_{SMR} \sin(2\theta_M) \frac{B_{DL}}{B_{ext}} \quad (11)$$

Equation 11 shows that, for the standard geometry discussed above, both the $B_{FL}$ (including the Oersted field) and $B_{DL}$ components acquire a $\sin(2\theta_M)$ dependence, which is well reproduced by the macrospin simulations. An accurate discrimination between the DL and FL-SOT signals at the $\varphi_B$ = 90º geometry is thus impossible by symmetry considerations. We note that the small difference in the field dependences are very difficult to discriminate due to low signal output levels typical of this system. Hence, the reported approaches typically assume that $B_{FL} \approx B_{DL}$.[4,46] On that basis, and considering that for most MIs $R_{SMR-AHE} \ll R_{SMR}$, the second harmonic resistance is approximated as arising solely due to the DL-SOT, leaving FL-SOT unquantified. The estimation of the FL-SOT can be normally achieved by using the other standard geometry $\varphi_B$ = 0º, where $B_{FL}$ drives the IP oscillations and its contribution to $R_{2\omega}^{H}$ is maximum. However, since the SSE contribution is proportional to $\cos\varphi_M$, its contribution dominates the second harmonic signal making the $B_{FL}$ quantification highly inaccurate, if not impossible.

In the above context, we suitably find the Hall signals at $\varphi_B$ = 85º and 95º as an alternative to circumvent the aforementioned issues. Indeed, deviations from $\varphi_B$ = 90º induce a significant asymmetry in the $B_{FL}$ contribution, thereby providing a convenient tool for its quantification. We note that this will occur at the expense of nonzero thermoelectric effects, so that a reasonable compromise is found at $\varphi_B$ = 85º and 95º for the parameter set considered in this study.

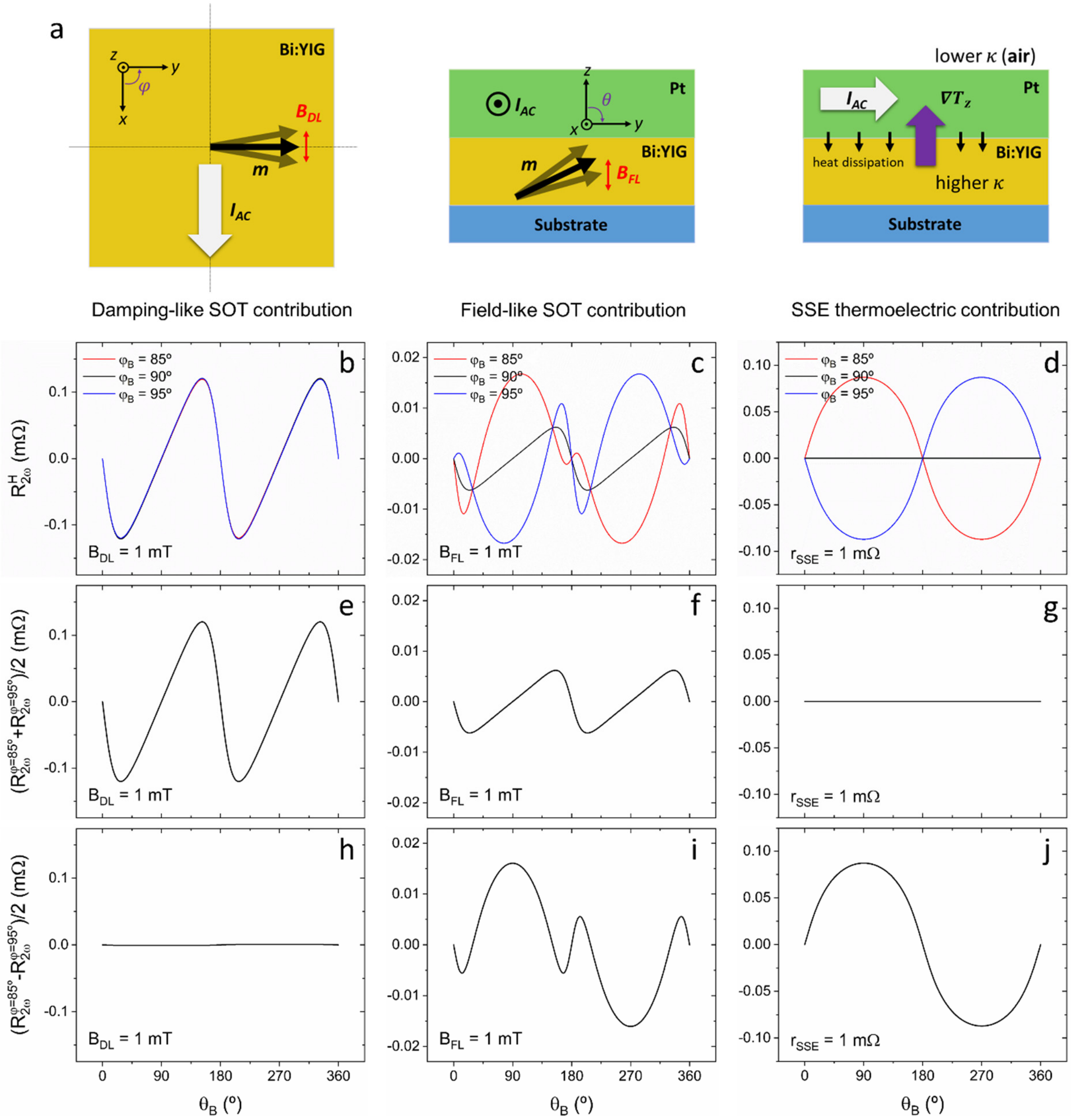

**Figure 3.** Symmetry-based toolbox herein proposed to disentangle the interplay between the DL, FL and SSE components. (a) Schematic representation depicting the symmetry of the current-induced fields at $\varphi_B$ = 90º geometry. Owing to their different symmetry with respect to $\varphi_B$ = 90º, the second harmonic oscillations originating from the previous individual contributions were simulated as a function of the azimuthal angle at $\varphi_B$ = 85º and 95º (top panel, b-d), and their average (middle panel, e-g) and difference (bottom panel, h-j) are herein proposed as reference signals for spin-orbit torque quantification.

## E. Separation of the DL-SOT, FL-SOT and thermoelectric signals

Figures 3b-d show simulated second harmonic $\theta$-scan curves at $\varphi_B$ = 85º, 90º and 95º (red, black, and blue lines). We discussed the standard case of $\varphi_B$ = 90º in the previous section and the problems associated with its consideration for the SOT quantification. For the $\varphi_B$ = 85º and 95º case, the first important observation is the nearly insensitivity of the DL-SOT component (Fig.3b) to small $\varphi_B$ variations around $\varphi_B$ = 90º. These changes account for 0.6% and will fall well below the detection limit in any given harmonic measurement. The contribution from the FL-SOT is, however, significantly distorted with respect to the $\varphi_B$ = 90º data and is amplified by a factor of 2.7. Moreover, and importantly, it has an opposite sign considering the $\theta_B$ ~90º and 270º data points as reference. Finally, the SSE contribution becomes now finite and has an opposite sign between $\varphi_B$ = 85º and 95º.

In an actual measurement, the signal will be a convolution of these three contributions. To separate the individual components, we proceed with the addition/subtraction of the transverse signals at $\varphi_B$ = 85º and 95º as the key step to disentangle the signals with different origins. The average (difference) of the second harmonic response at these two $\varphi_B$ angles, hereafter denoted as $R_{2\omega}^{85+95}$ ($R_{2\omega}^{85-95}$) for simplicity, is depicted in Figures 3e-g (Figures 3h-j). As expected, $R_{2\omega}^{85+95}$ exhibits the same symmetry and magnitude as in the $\varphi_B$ = 90º case for all three components, with a calculated variation as small as 0.6% and 0.3% for the DL and FL-SOT contributions. On the other hand, $R_{2\omega}^{85-95}$ effectively suppresses the contribution from the DL-SOT, resulting in a combined response of the FL-SOT and SSE effects. It can be observed that $R_{2\omega}^{85-95}$ displays a maximum (minimum) at $\theta_B$ = 90º (270º) for the latter two components. Note that in the SSE case it is a direct consequence of the $\sin\theta_M$ dependence as $\theta_M \approx \theta_B$ holds throughout the entire scan. Further separation between the FL-SOT and SSE can be achieved analyzing the $R_{2\omega}^{85-95}$ variation with the external field amplitude. For increasing $B_{ext}$, the amplitude of the $B_{FL}$-induced oscillations are reduced, as the magnetization becomes more strongly coupled to the field direction. The SSE constitutes instead a static effect, which depends on the magnetization orientation but not on the field amplitude.[17] The exact external field dependence of $R_{2\omega}^{85-95}$ at $\theta_B$ = 90º can be ascertained from the second harmonic analysis. In the vicinity of $\theta_B$ = 90º, eq. 7 reduces to:

$$R_{2\omega}^{H} = R_{SMR-AHE}\frac{d\cos\theta_M}{d\theta_M}\frac{B_I^{\theta}}{B_{ext}+B_{dem}} + R_{SMR}\frac{d\sin(2\varphi_M)}{d\varphi_M}\frac{B_I^{\varphi}}{B_{ext}} + r_{SSE}\cos\varphi_M \quad (12)$$

At this $\theta_B$ angle, the angular components of the current-induced fields in eq. 12 are given by $B_I^{\theta}$ = $\cos\varphi_M B_{DL}$ and $B_I^{\varphi}$ = $\cos\varphi_M B_{FL}$. Upon calculating the derivative terms, we obtain:

$$R_{2\omega}^{H} = -R_{SMR-AHE}\cos\varphi_M\frac{B_{DL}}{B_{ext}+B_{dem}} + 2R_{SMR}\cos(2\varphi_M)\cos\varphi_M\frac{B_{FL+Oe}}{B_{ext}} + r_{SSE}\cos\varphi_M$$
$$= R_{2\omega}^{DL} + R_{2\omega}^{FL} + R_{2\omega}^{SSE} \quad (13)$$

Note that the contributions of the DL-SOT, FL-SOT and SSE to the transverse second harmonic resistance ($R_{2\omega}^{DL}$, $R_{2\omega}^{FL}$ and $R_{2\omega}^{SSE}$, respectively) appear as separate terms. Summarizing these relations, and taking into account that $\cos(90° - x) = -\cos(90° + x)$, the following expression is derived for the $R_{2\omega}^{85-95}$ parameter:

$$R_{2\omega}^{85-95} = -R_{SMR-AHE}\frac{B_{DL}}{B_{ext} + B_{dem}}\cos 85° + 2R_{SMR}\frac{B_{FL+Oe}}{B_{ext}}(2\cos^3 85° - \cos 85°) + r_{SSE}\cos 85° \quad (14)$$

The discrimination between the DL and FL-SOT can be then accomplished through their different dependencies on $B_{ext}$. In fact, since $R_{SMR-AHE} \ll 2R_{SMR}$ and the effective field acting against $R_{2\omega}^{DL}$ is larger than that of $R_{2\omega}^{FL}$ because of $B_{dem}$, the second term in eq. 14 significantly dominates over the first one. This leads to a linear relationship between $R_{2\omega}^{85-95}$ and $1/B_{ext}$, according to which the slope is modulated by the product of $R_{SMR}$ and $B_{FL}$, and $r_{SSE}$ provides a constant offset independent of the external field.

We verified the above calculations by means of macrospin simulations. Figure 4a summarizes the magnetic field dependence of $R_{2\omega}^{85-95}$, simulated with parameters of $B_{DL}$ = $B_{FL}$ = 1 mT and $r_{SSE}$ = 1 mΩ. For a better appreciation, a close-up view of the same plot about $\theta_B$ = 90º is displayed in Figure 4b. The amplitude modulation of the peak at $\theta_B$ = 90º with $B_{ext}$ is predominantly due to the FL-SOT. The SSE contribution to the data is constant at $\theta_B$ = 90º or 270º and only the width of the peak changes due to stronger coupling of the magnetization vector with the external field at higher $B_{ext}$ (see supplementary Figure S1c). Exploiting this particular behavior, the peak amplitude is plotted as a function of the inverse of the external field in Figure 4c. In agreement with eq. 14 the data can be fitted to a straight line and $B_{FL}$ can be extracted by dividing the slope by $2R_{SMR}(2\cos^3 85° - \cos 85°)$. The y-axis intercept of the linear fitting yields $r_{SSE}$ upon normalizing over $\cos 85°$. The quantification of $B_{DL}$ can be finally accomplished through the $R_{2\omega}^{85+95}$ data or the second harmonic resistance measured at $\varphi_B$ = 90º ($R_{2\omega}^{90}$). In both data the $r_{SSE}$ contribution vanishes, such that $B_{DL}$ can then be deduced by quantitatively comparing the macrospin simulations (performed with fixed $B_{FL}$ and variable $B_{DL}$ parameters) with the experimental data. The as-described methodology will hold for moderate $B_{DL}/B_{FL}$ ratios, as demonstrated by the error estimation in these two parameters (included in Figure S1e). Nevertheless, when $B_{DL}$>>$B_{FL}$ a strong non-linearity will be observed in the $R_{2\omega}^{85-95}$ vs. $1/B_{ext}$ data, leading to an increasing systematic error in the determination of $B_{FL}$. For such scenario an iterative approach should be followed, so that the calculated $B_{DL}$ is recursively used as an input parameter for $B_{FL}$ quantification via fitting of the $R_{2\omega}^{85-95}$ data according to eq. 14. This procedure allows for an intrinsic error in $B_{FL}$ of only 0.6% even when considering the unfavorable scenario in which $B_{DL}/B_{FL}$=10 (refer to Figure S1f for estimation errors upon conducting the previous iterative analysis).

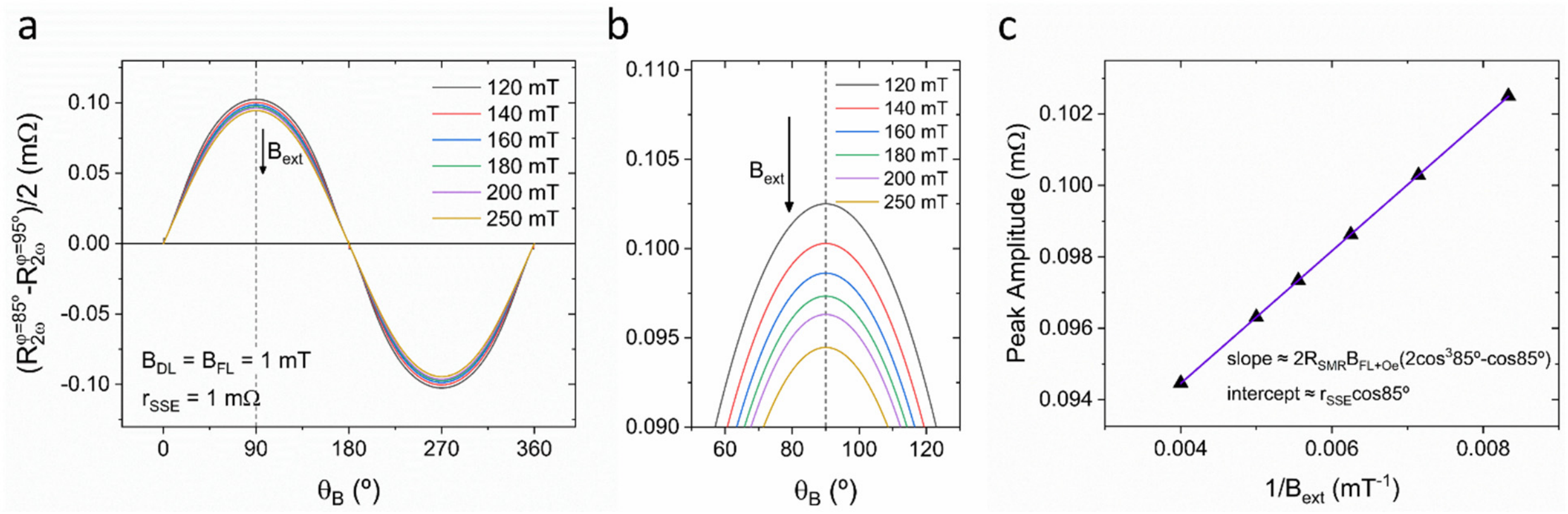

**Figure 4.** (a) External field dependence of $R_{2\omega}^{85-95}$ and (b) a close-up view of the same plot around $\theta$= 90°; (c) $R_{2\omega}^{85-95}$ (measured at $\theta$= 90°) as a function of the inverse external field. A linear fit to this data allows for the quantification of $B_{FL}$ and $r_{SSE}$ by dividing the slope and intercept over $2R_{SMR}(2\cos^3 85° - \cos 85°)$ and $\cos 85°$, respectively.

### F. Experimental results and discussion

#### a. *Second harmonic measurements of SOTs in Bi:YIG/Pt*

As a proof-of-concept, the described methodology has been applied to the Bi:YIG/Pt device described in Section III-A. The second harmonic Hall resistance was measured while sweeping the polar field angle $\theta_B$ for a fixed magnitude of 120 mT and a current density of $j$ = 1.7×$10^{11}$ A/$m^2$. The measurements were performed at a $\varphi_B$ angle of 85º or 95º, as shown in Figure 5a (in red and blue, respectively). The lineshape of the as-measured signals greatly differs from that simulated in Figure 3. In fact, from the simulations it can be seen that all current-induced effects are antisymmetric with respect to $\theta_B$ = 180º. This distinctive property allows us to identify and separate relevant SOT and SSE signals from other spurious signals based on symmetry operations. Figures 5b,c show the antisymmetric and symmetric components of the raw data where only the former is considered for the HHV analysis. The lineshape of the additional (spurious) symmetric component resembles that of the first harmonic signal originating from $R_{SMR-AHE}$ and $R_{OHE}$. Therefore, we believe that the signal in the second harmonic is related to an in-plane temperature gradient in the device along the current injection line, producing the thermal counterparts of $R_{SMR-AHE}$ and $R_{OHE}$.

The external field dependence of $R_{2\omega}^{85-95}$, determined from the experimental data in Figure 5b, is displayed in Figure 5d. Notice that the sign of $R_{2\omega}^{85-95}$ is opposite to the simulations and that the $\theta_B$ = 90º peak amplitude becomes less negative as increasing $B_{ext}$. This is consistent with negative $r_{SSE}$ and positive $B_{FL+Oe}$ parameters. The dependence of this peak amplitude as a function of 1/$B_{ext}$ is plotted in Figure 5e, whose linear fitting reveals $r_{SSE}$ and $B_{FL+Oe}$ values of -0.39(±0.01) mΩ and 0.29(±0.04) mT respectively. Upon subtraction of the Oersted field, which is assumed to be linearly proportional to the current ($B_{Oe}$ = 0.43 mT), we obtain $B_{FL}$ = -0.14($\pm$0.04) mT. The quantification of $B_{DL}$ was then addressed by measuring the second harmonic resistance at $\varphi_B$ = 90º (i.e. $R_{2\omega}^{90}$, see Figure 5f). Rescaling the simulated $\theta$-scans for the DL and FL components (shown in Figures 3b,c respectively) with the as-calculated $B_{FL}$ and variable $B_{DL}$ parameters, a quantitative comparison of the simulations with the experimental data reveals a $B_{DL}$ value of -0.15 mT. This result was also verified with the $R_{2\omega}^{85+95}$ data (not shown). Ignoring $B_{FL}$ instead results in a non-negligible overestimation of $B_{DL}$ of about 6.7%, proving the necessity and relevance of the hereby described method.

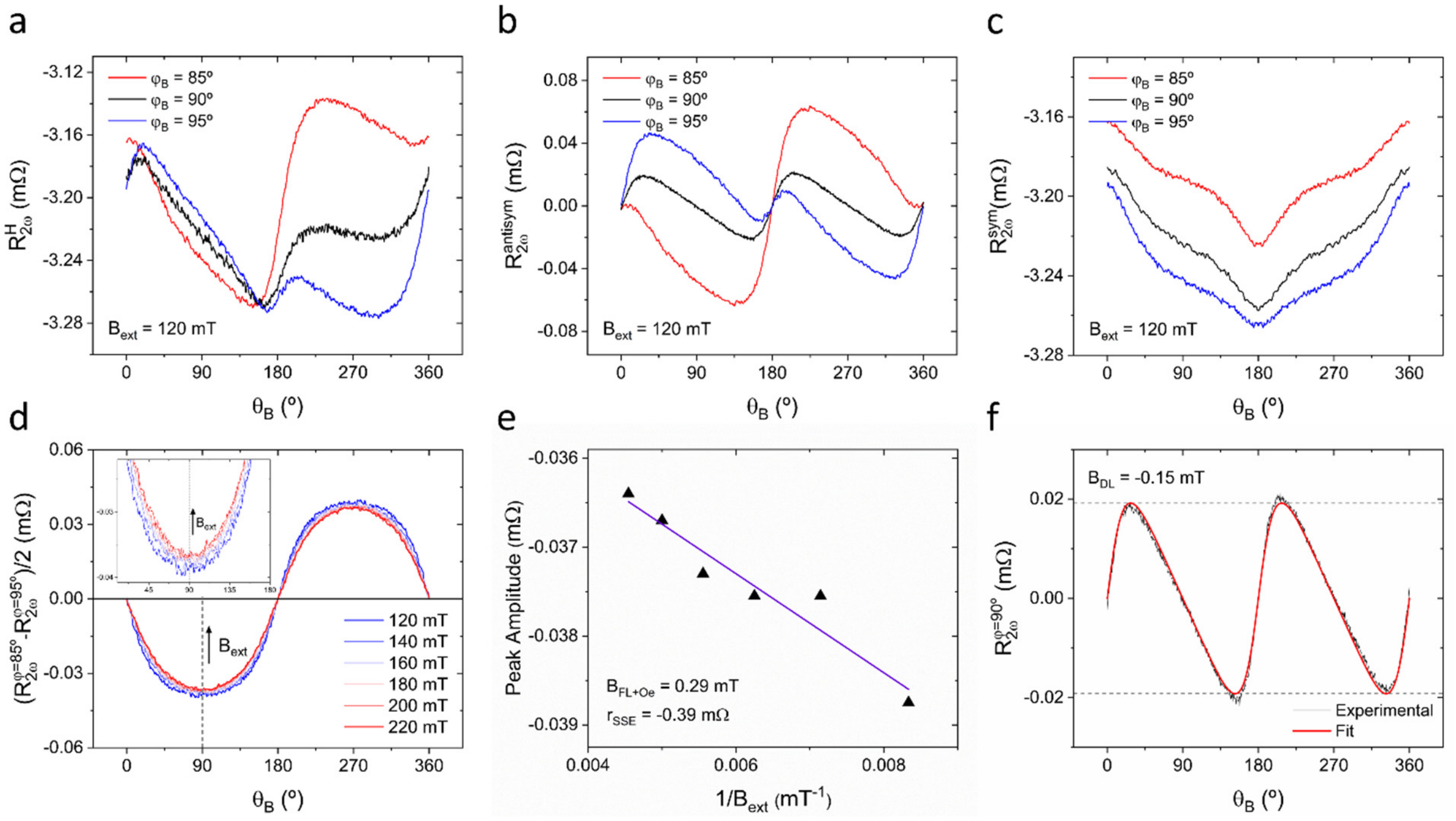

**Figure 5.** Quantification of the SOTs via second harmonic Hall measurements: (a) Raw second harmonic transverse resistance as a function of the azimuthal angle (at $\varphi_B$ = 85°, 90° and 95°), applying an external field of 120 mT; (b) Antisymmetric and (c) symmetric components of the data shown in **a**, being the former hereafter employed for further quantitative analysis; (d) Variation of the $R_{2\omega}^{85-95}$ signal and (e) its amplitude (at $\theta_B$= 90°) with the inverse of the external field. (f) Fitting of the $R_{2\omega}^{H}$ response at $\varphi_B$ = 90° taking into account the damping-like and field-like contributions simulated in Figures 3b,c. All measurements were conducted at a current density (rms) of $j$ = 1.7×10$^{11}$ A/m$^2$.

The values of $B_{DL}$ and $B_{FL}$ reported in Section III-F,a are significantly lower than those reported in metallic films[16,17,69] and in the specific case of $B_{DL}$, it is somewhat lower than those found in other MI/Pt systems[4,46]. For quantitative comparison, we convert $B_{DL}$ into an effective spin Hall angle (SHA) assuming the SHE as the sole spin current generating mechanism:[70]

$$\theta_{SH} = \frac{2e}{\hbar}\frac{M_s t_{Bi:YIG} B_{DL}}{j} \quad (15)$$

where $e$ stands for the electron charge, $\hbar$ the reduced Planck constant, and $M_s$ and $t_{Bi:YIG}$ the saturation magnetization and thickness of the Bi:YIG layer. We note that eq.15 does not take into account the effect of spin diffusion length and assume full absorption of the spin current without cancelation effect due to the opposite spin accumulation at the counter interface. We find a SHA value of ~0.4%, which is lower than the common values reported for YIG/Pt bilayers.[35] Assuming that Pt deposited in our chamber has comparable bulk properties to the systems reported in literature, a relatively small effective SHA in our system can have multiple interfacial origins. One common difference is due to the inefficiency of the conversion of the spin current into damping-like spin-torque. This can be due to a low spin mixing conductance or large spin memory loss[71,72] at the Bi:YIG/Pt interface. Another potential origin is related to the density of magnetic ions at the interface. It has been speculated that a reduced magnetization would also reduce the capability of the magnetic layer to absorb the spin current and convert it into spin-torque.[48,73,74] Due to $Bi^{3+}$ doping and consequently low $M_s$ value, it is plausible that we obtain a lower SHA with respect to other garnet systems studied thus far.

Another remarkable observation is that the magnitude of $B_{FL}$ is comparable to that of $B_{DL}$. Typically, the damping-like and field-like torques are associated with the real ($G_r^{\uparrow\downarrow}$) and imaginary ($G_i^{\uparrow\downarrow}$) parts of the spin mixing conductances, respectively. In this picture, judging from the $R_{SMR}$ (proportional to $G_r^{\uparrow\downarrow}$) and $R_{SMR-AHE}$ (proportional to $G_i^{\uparrow\downarrow}$), $B_{DL}$ should be about a factor 8 larger than $B_{FL}$. The large $B_{FL}$ in our system suggests that, either this commonly accepted picture should be revised (if SHE is assumed to be the sole SOT source) or some additional SOT-generating mechanisms exist at the Bi:YIG/Pt interface. Plausible mechanisms include the occurrence of the Rashba-Edelstein effect generating additional SOTs with a stronger field-like contribution or magnetic proximity effect in Pt acting as an additional source of spin scattering near the interface favoring field-like SOT generation.[75] Nevertheless, investigating the SOT anomalies and deviations with respect to the literature are beyond the scope of the present work.

### *b. Current dependence of SOTs in Bi:YIG/Pt*

In typical metallic SOTs systems (e.g., Pt/Co), current-induced Joule heating plays a minor role in the first and second harmonic responses, hence generally neglected in measurements with moderate current densities ($j$ < 2×10$^{11}$ A/m$^2$).[16] It is typically assumed that the parameters giving rise to the first harmonic response are independent of the current, meanwhile the SOTs and thermoelectric effects (expressed in resistance) depends linearly in current. However, in MI/Pt, comparable current densities can create a larger Joule heating due to lower thermal conductivity of the MI and the substrate, which can cause a significant modulation to the SMR parameters through the temperature-dependence of interfacial spin mixing conductance,[76] as well as a decrease of the magnetization, magnetocrystalline or magnetoelastic anisotropies.[77]

In Figure 6a, we report the current-dependence of the $R_{SMR}$, $R_{SMR-AHE}$ and $B_{DL}$ in Bi:YIG/Pt in the current range of $j$ = 0.5-2×10[11] A/m[2]. Assuming that the first data point at $j$ =0.5×10[11] A/m[2] is representative of room temperature values, we find that the magnitude of $R_{SMR}$ progressively decreases by about 16% when increasing the current density to 2×10[11] A/m[2], whereas $R_{SMR-AHE}$ increases by 43% in the same current range. Surprisingly, $B_{DL}$ departs from its expected linear trend at about $j$ = 1.25×10[11] A/m[2] and follows a decreasing tendency at elevated current densities. At first approximation, $R_{SMR}$ and $B_{DL}$ are both directly related to the real part of the spin-mixing conductance of the Bi:YIG/Pt interface. In order to compare them on equal grounds, we subtract a linear background from the $B_{DL}$ values using the data points corresponding to $j$ ≤ 1.0×10[11] A/m[2] and then normalize both $B_{DL}$ and $R_{SMR}$ dataset by dividing them by the value obtained at the lowest current density, i.e., $j$ = 0.5×10[11] A/m[2]. Figure 6c shows the comparison of the normalized data displaying a strong correlation between these two sets of data and pinpointing their expected common origin.

These results collectively show that the spin-dependent parameters of Bi:YIG/Pt critically depend on the measurement temperature even for modest current densities. This behavior is tentatively attributed to the strong sensitivity of magnetic and spin-dependent interfacial properties of Bi:YIG to temperature variations created by Joule heating. Especially, in light of the literature,[78,79] we believe that the decreasing spin-mixing conductance upon increasing temperature is the likely cause of the reduced damping-like SOT efficiency and $R_{SMR}$ collectively.

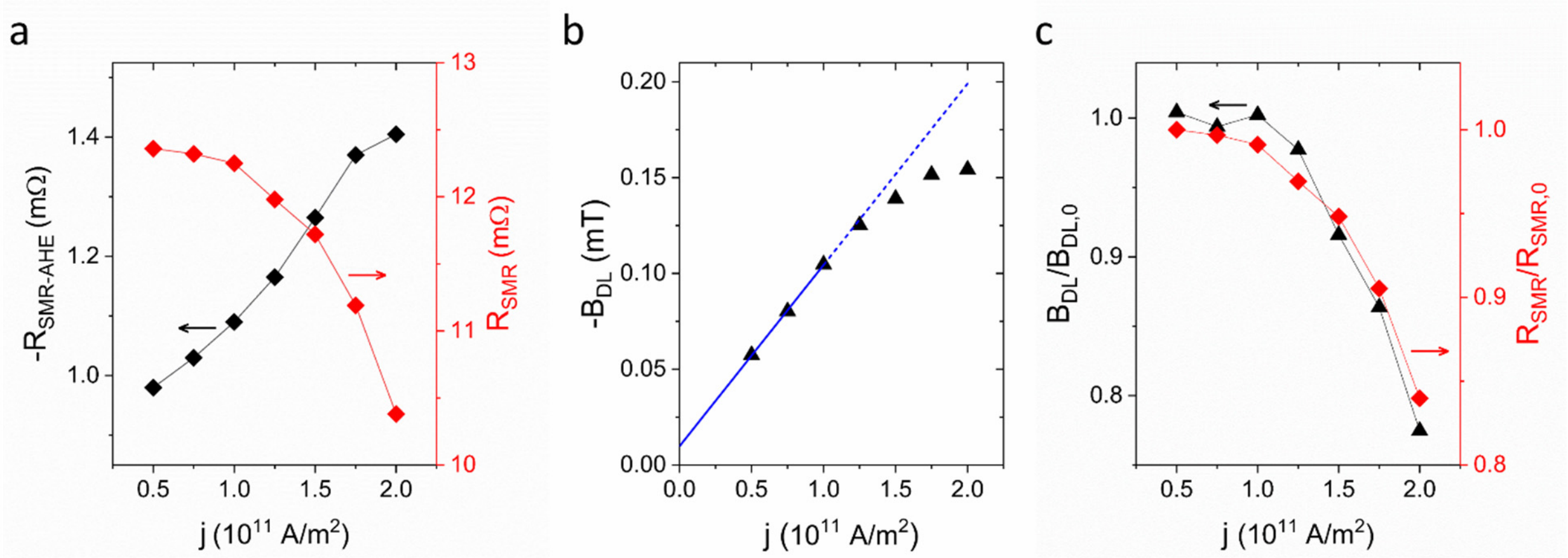

**Figure 6.** (a) Modulation of the $R_{SMR-AHE}$ and $R_{SMR}$ parameters as a function of the current density (in rms) due to Joule heating and (b) as-derived current dependence of $B_{DL}$. A small deviation from linearity is observed at large $j$, even after considering the current-corrected values of $R_{SMR-AHE}$ and $R_{SMR}$, which highlights a non-negligible temperature-dependence of $B_{DL}$. Note that the linear fitting in **b** has been performed considering only the experimental data for $j$ ≤ 1.0×10[11] A/m[2]. (c) Correlation between the current

modulation of the $R_{SMR}$ coefficient and the linear deviation in $B_{DL}$. For that, the $R_{SMR}$ and $B_{DL}$ parameters were normalized over the experimental values obtained at $j$ = 0.5×10$^{11}$ A/m$^2$ ($R_{SMR,0}$ and $B_{DL,0}$, respectively), where no moderate Joule heating is expected.

## IV. CONCLUSIONS

In summary, the study herein presented offers an alternative scheme for SOT vector characterization in MIs by shifting to a non-standard geometry at $\varphi_B$ = 85º and 95º. Supported by macrospin simulations and symmetry arguments, we have developed a consistent methodology that allows to accurately quantify the damping-like and field-like SOTs and thermoelectric contributions separately, with an inherent error on the simulated data well below 1% for all three components. As a proof-of-concept, this strategy was tested on an IP-magnetized Bi:YIG/Pt heterostructure, obtaining $B_{DL}$ and $B_{FL}$ values of -0.15 and -0.14 mT per $j$ = 1.7×10$^{11}$ A/m$^2$ injected current, respectively. Despite the appropriateness of the devised method for this sample, we further show how a recursive procedure can be equally applied without compromising the accuracy of the resultant data to systems displaying higher $B_{DL}/B_{FL}$ ratios, where the $B_{FL}$ quantification becomes remarkedly challenging and inaccurate. This helps broaden its applicability to a wide range of materials and confirms the robustness of the proposed methodology, also found to be stable against moderate azimuthal angle misalignments. Finally, we demonstrate that Joule heating can modulate the spin transparency at the interface, even for modest current densities. This leads to a systematic reduction of the SMR and the DL-SOT, which is often neglected in most studies up to date. Overall, this study reinforces the need of an adequate SOT characterization, which is of paramount importance for the design of MI-based spin Hall nano-oscillators and novel magnonics devices, among others.

## ACKNOWLEDGEMENTS

The authors acknowledge funding from the European Research Council (ERC) under the European Union's Horizon 2020 research and innovation programme (project MAGNEPIC, grant agreement No. 949052) and from the Spanish Ministry of Science and Innovation through grant reference No. PID2021-125973OA-I00. M. T.-A. acknowledges financial support from the Spanish Ministry of Science and Innovation under grant FJC2021-046680-I.

## CONFLICT OF INTEREST

The authors declare no conflict of interest.

# Supporting Information

## Leveraging symmetry for an accurate spin-orbit torques characterization in ferrimagnetic insulators

Martín Testa-Anta[1,*], Charles-Henri Lambert[2], Can Onur Avci[1,*]

[1]*Institut de Ciència de Materials de Barcelona (ICMAB-CSIC), Campus de la UAB, 08193 Bellaterra, Spain*

[2]*Department of Materials, ETH Zürich, Hönggerbergring 64, CH-8093 Zürich, Switzerland*

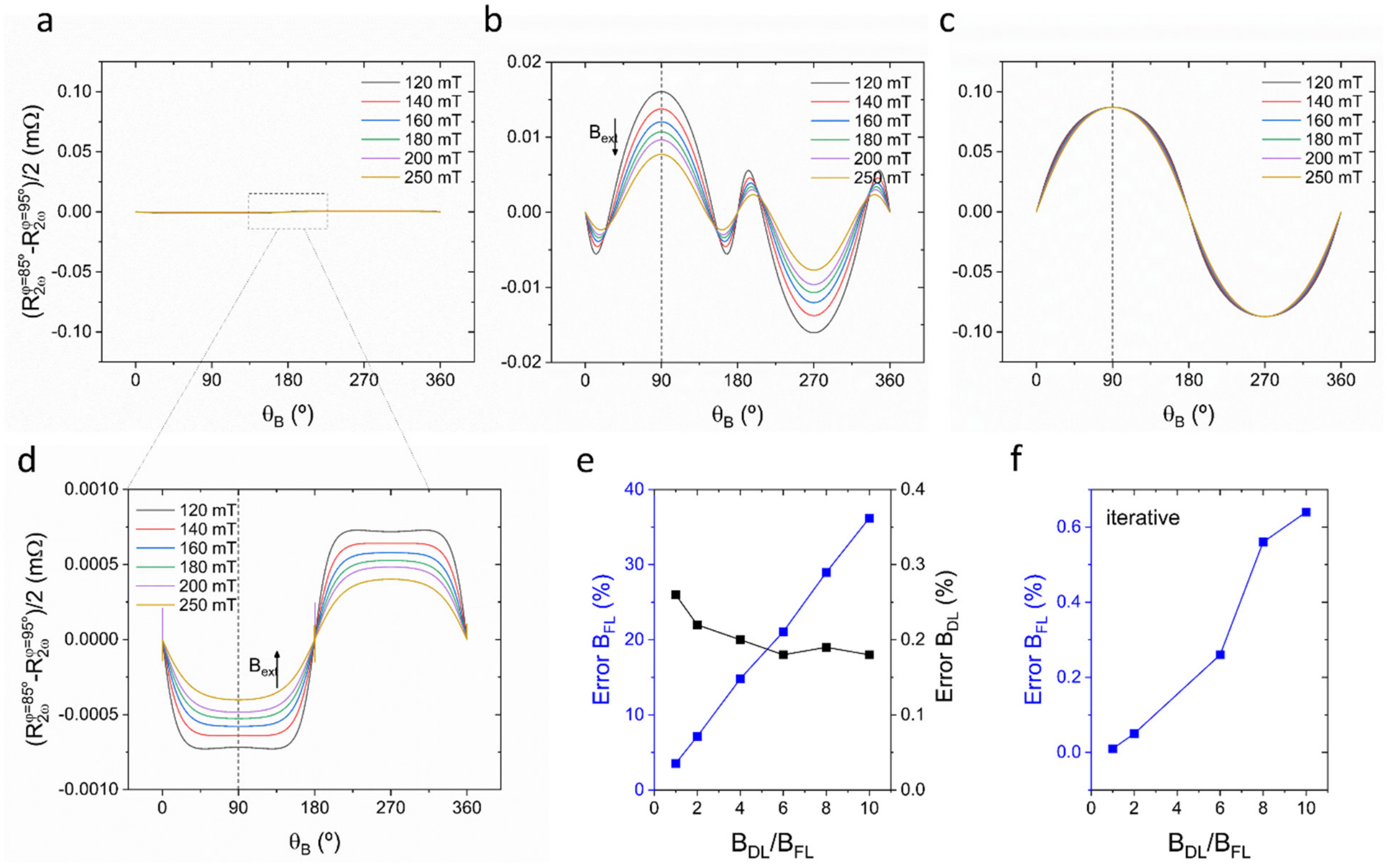

**Figure S1**. External field dependence of $R_{2\omega}^{85-95}$ considering the individual (a) DL, (b) FL and (c) SSE contributions; (d) Zoom-out of the same plot shown in **a**, displaying the residual contribution to $R_{2\omega}^{85-95}$ due to the OOP oscillations induced by $B_{DL}$. This contribution was neglected in the SOT analysis, as being ~20 and ~120 times smaller than those of the FL-SOT and SSE, respectively. (e) Estimated errors in the field-like and damping-like fields for different $B_{DL}/B_{FL}$ ratios owing to the contribution shown in **d**. Considering a system where $B_{DL}$ = $B_{FL}$ = 1 mT and $r_{SSE}$ = 1 mΩ, the estimated error is calculated to be 0.26, 3.54 and 0.13% for $B_{DL}$, $B_{FL}$ and $r_{SSE}$ respectively, which is expected to be hidden within the experimental noise. Note that the error in $B_{DL}$ ($B_{FL}$) will slightly decrease (significantly increase) as increasing the $B_{DL}/B_{FL}$ ratio, as depicted in **e**. For such scenario an iterative process is proposed, in which the calculated $B_{DL}$ is used as an input parameter for the estimation of $B_{FL}$ via fitting of the $R_{2\omega}^{85-95}$ vs. $B_{ext}$ experimental data (according to eq. 14 in the manuscript). The remarked reduction of the estimated error in the $B_{FL}$ parameter after conducting this iterative process is shown in **f**.

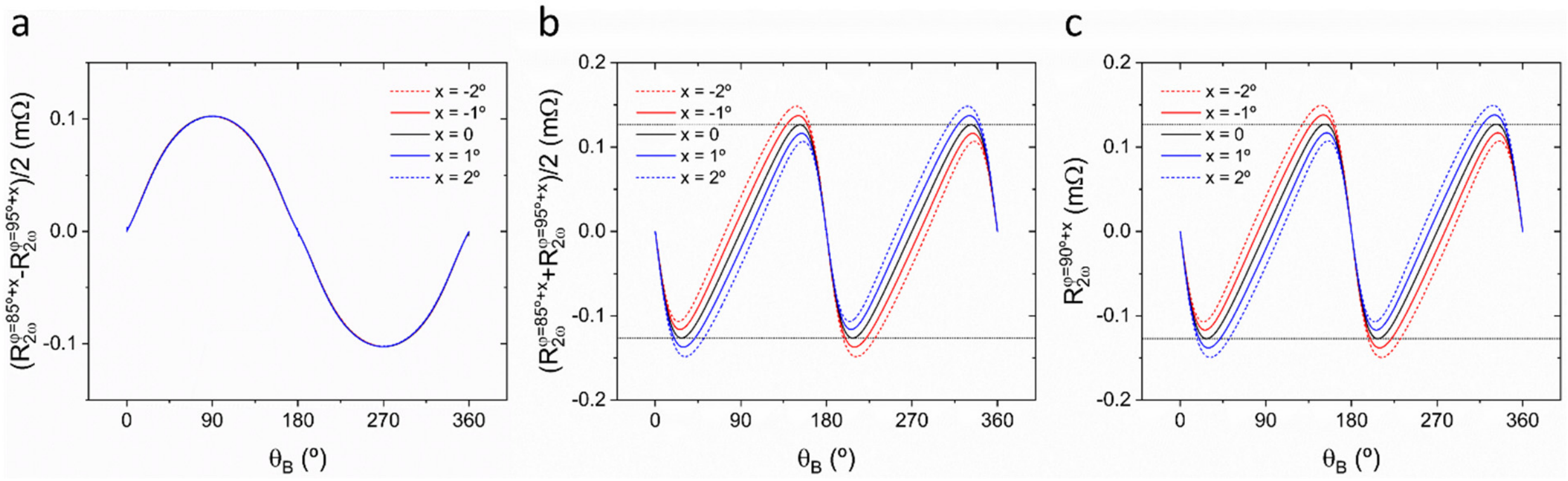

**Figure S2.** (a) $R_{2\omega}^{85-95}$, (b) $R_{2\omega}^{85+95}$ and (c) $R_{2\omega}^{90}$ as a function of the azimuthal angle for different $\varphi_B$-angle off-centerings. A critical factor when registering the second harmonic response is the Hall bar alignment with respect to the external field. Owing to the large SSE intrinsic to the Bi:YIG layer, small off-centerings from $\varphi_B$ = 90º (even <<1º) will lead to a significant SSE contribution that will manifest through changes in the relative intensity of the peaks at $\theta \sim$ 28º and 208º, as shown in **c**. An identical parasitic $\sin\theta_M$ contribution is observed in the $R_{2\omega}^{85+95}$ response for different off-centerings (see graph **b**), provided that the measurements at the two $\varphi_B$ angles span a constant 10º range. In any case, in light of the antisymmetric behavior of the SOTs with respect to $\theta$ = 180º, the experimental occurrence of a $\varphi_B$ off-centering can be corrected in both $R_{2\omega}^{90}$ and $R_{2\omega}^{85+95}$ by taking the average of the peaks at $\theta \sim$ 28º and 208º as the actual amplitude of the second harmonic resistance. This average is indicated by horizontal dotted lines in **b** and **c**. Alternatively, $R_{2\omega}^{85-95}$ sees a negligible influence on a potential off-centering (see graph **a**), meaning that neither $B_{FL}$ nor $r_{SSE}$ will be affected by the misalignment.

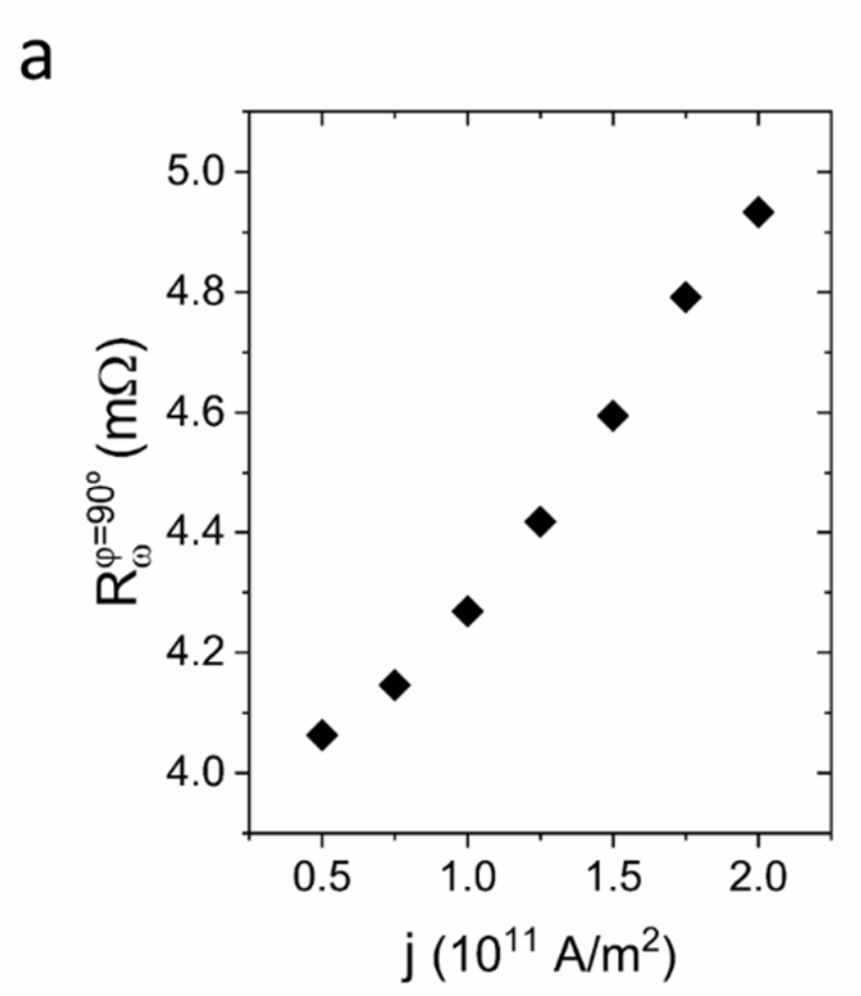

**Figure S3.** Current-dependence of the (a) first and (b) second harmonic resistance measured at $\varphi_B$ = 90º (under a 120 mT external field). At this configuration, the second harmonic resistance encompasses the DL and FL-SOT contributions, without the influence of the thermoelectric SSE effect. A remarked deviation from linearity can be observed at high $j$, which stems from the Joule heating modulation of $R_{SMR}$ and, subsequently, of the in-plane oscillations.

## SciLab script for macrospin simulations

```
clear;
function T_tot=T_min(angle)

    theta_M=angle(1);
    phi_M=angle(2);
    zeta_M=angle(3);

//---Current-induced fields scaling with the injected current---------------

H_FL=I_DC*HFL;
H_DL=I_DC*HDL;

//---Torques acting on the magnetization---------------

T_ext=[H*M_s*(sin(theta_M)*sin(phi_M)*cos(theta_H(ii))-cos(theta_M)*sin(phi_H)*sin(theta_H(ii)));
            H*M_s*(-sin(theta_M)*cos(phi_M)*cos(theta_H(ii))+cos(theta_M)*cos(phi_H)*sin(theta_H(ii)));
            H*M_s*(sin(theta_M)*cos(phi_M)*sin(phi_H)*sin(theta_H(ii))-sin(theta_M)*sin(phi_M)*sin(theta_H(ii))*cos(phi_H))];

T_dem=[-0.5*H_dem*M_s*sin(2*theta_M)*sin(phi_M);
            0.5*H_dem*M_s*sin(2*theta_M)*cos(phi_M);
            H_dem*M_s*0];

T_FL=[H_FL*M_s*(cos(theta_M));
         H_FL*M_s*0;
        -H_FL*M_s*sin(theta_M)*cos(phi_M)]

T_DL=[-H_DL*sin(theta_M)^2*sin(phi_M)*cos(phi_M);
          H_DL*(cos(theta_M)^2+sin(theta_M)^2*cos(phi_M)^2);
         -H_DL*cos(theta_M)*sin(theta_M)*sin(phi_M)];

    T_tot(1)=T_ext(1)+T_dem(1)+T_FL(1)+T_DL(1);
    T_tot(2)=T_ext(2)+T_dem(2)+T_FL(2)+T_DL(2);
    T_tot(3)=T_ext(3)+T_dem(3)+T_FL(3)+T_DL(3);

endfunction

//---External parameters---------------

I_app=1;                        //Scaling parameter related to the applied current
theta_Hdeg=[0:0.1:360];         //Out-of-plane field angle with respect to z-axis
phi_H=95*%pi/180;               //In-plane field angle with respect to current injection direction
Hext=[120];                     //External field amplitude (in mT)

//---Magnetic, electrical and thermal parameters---------------

H_dem=45;                       //Demagnetizing field (in mT)
M_s=1;                          //Scaling parameter related to the saturation magnetization
SMR=11.23;                      //Spin Hall magnetoresistance (in mOhm)
SMR_AHE=-1.36;                  //Anomalous Hall-like spin Hall magnetoresistance (in mOhm)
HFL=1;                          //Field-like SOT effective field (in mT)
HDL=1;                          //Damping-like SOT effective field (in mT)
r_SSE=1;                        //SSE coefficient (in mOhm) for a temperature gradient along z-axis
```

```
//---Data initialization---------------

theta_H=[];

R_xy1=[];               R_xy2=[];          //Transverse resistance for I+ and I-
Rth_xy1=[];             Rth_xy2=[];        //Thermal contribution to transverse resistance for I+ and I-

Ph1=[];                 Ph2=[];            //Equilibrium phi_M
Th1=[];                 Th2=[];            //Equilibrium theta_M

R_xy=[];                                   //First harmonic transverse resistance
Delta_R_xy=[];                             //Second harmonic transverse resistance

init1=theta_Hdeg(1)*%pi/180;               //Initial value of theta_M
init2=phi_H;                               //initial value of phi_M

//---Computation---------------

for hh=1:length(Hext)                      //Loop for different external field amplitude values
H=Hext(hh);

for ii=1:length(theta_Hdeg);               //Loop for different external field theta angle values
theta_H(ii)=theta_Hdeg(ii)*%pi/180;

//--- Computation for positive current I+ ---------------

I_DC=+I_app;

for jj=0:10;
        [t]=fsolve([init1;init2;0],T_min,1e-15);
        init1=t(1); init2=t(2);
end

Ph1(ii)=2*t(2);
Th1(ii)=t(1);

Rth_xy1(ii)= r_SSE*sin(t(1))*cos(t(2));

R_xy1(ii)=I_app*(SMR*(sin(t(1))^2)*sin(2*t(2)))+ SMR_AHE *cos(t(1))+Rth_xy1(ii);

//--- Computation for negative current I- ---------------

I_DC=-I_app;

for kk=0:10;
        [t]=fsolve([init1;init2;0],T_min,1e-15);
        init1=t(1); init2=t(2);
end

Ph2(ii)=2*t(2);
Th2(ii)=t(1);

Rth_xy2(ii)=-(r_SSE*sin(t(1))*cos(t(2)));

R_xy2(ii)=I_app*(SMR*(sin(t(1))^2)*sin(2*t(2)))+SMR_AHE*cos(t(1))+Rth_xy2(ii);
```

```
//--- Computing the average and the difference between I+ and I- ---------------

R_xy(ii)=(R_xy1(ii)+R_xy2(ii))/2;                //Results for the first harmonic transverse resistance
Delta_R_xy(ii)=(R_xy1(ii)-R_xy2(ii))/2;          //Results for the second harmonic transverse resistance

end

//---Data saving - Output---------------

out=[theta_Hdeg' R_xy Delta_R_xy];
writefile='Data saving directory\filename.dat'
fprintfMat(writefile,out,'%.10g');

end
```